\documentclass[12pt,a4paper]{article}

\usepackage{amsmath,amssymb}

\begin{document}

\bibliographystyle{unsrt}

\title{Raising and lowering operators and their factorization for generalized orthogonal polynomials of
hypergeometric type on homogeneous and non-homogeneous lattice}

\author{Miguel Lorente\\
\small Departamento de F\'{\i}sica, Universidad de Oviedo, 33007 Oviedo, Spain}
\date{}
\maketitle

\begin{abstract} We complete the construction of raising and lowering operators, given in a previous work, for the
orthogonal polynomials of hypergeometric type on non-homogeneous lattice, and extend these operators to the
generalized orthogonal polynomials, namely, those difference of orthogonal polynomials that satisfy a similar
difference equation of hypergeometric type.

\end{abstract}

\section{Introduction}

Recently we have presented a paper on the raising and lowering operators for the orthogo\-nal polynomials (OP) of
hypergeometric type [5] (in connection with the factorization method defined by Hull and Infeld). That paper covered
only OP of continuous and discrete variable on an uniform lattice, as well as orthonormal functions of continuous
and discrete variable.

In this work we continue the construction of raising and lowering operators for OP on non-homogeneous lattice. The
starting point is also the Rodriguez formula and the fundamental properties of OP given by Nikiforov and
collaborators [1], [18] that include the q-analog of classical OP of discrete variable.

Atakishiyev and collaborators [13] [14] extended the classification of Nikiforov to OP of discrete variable defined
by Andrews and Askey and proved that they satisfy a difference equation only in the case of $x(s)$ linear, quadratic
q-linear and q-quadratic.

The construction of raising and lowering operators on non-uniform lattice was also worked out by Alvarez-Nodarse and
Costas-Santos [15][16] for the lattice $x(s)=c_1q^s+c_2q^{-s}+c_3.$

Similar work was carried out by Smirnov for OP of hypergeometric type on homogeneous lattice [20] and
non-homogeneous lattice [21]although the raising and lowering operators are defined with respect to two indices:
$n$, the order of polinomials {\bf and} $m$, the order of difference derivative of polynomials. In our work we
define the raising and lowering operator with respect to one index only, $n$ {\bf or} $m$.

In order to complete the classification of the OP of hypergeometric type we include the generalized classical OP that
satisfy also a difference/differential equation of hypergeo\-metric type.

Since all classical OP of discrete variable lead in the limit to the corresponding OP of continuous variable, we
start in section 2 with the raising and lowering operators for generalized OP of continuous variable with respect to
the index $n$, using Rodrigues formula. In section 3 we repeat the same construction for generalized classical OP of
discrete variable on hmogeneous lattice. In section 4 we extend the construction to classical OP of discrete
variable on non-homogeneous lattice in the general case, when $x(s)=c_1s^2+c_2s+c_3$ or
$x(s)=c_1q^s+c_2q^{-s}+c_3$.

In section 5 we complete the picture with the construction of raising and lowering ope\-rators for generalized
classical OP on non-homogeneous lattice, that include the q-analog of classical OP of discrete variable. In all
these cases, the raising and lowering operators are given with respect to one index, say, $n$, but the same operator
can be considered, written in appropiate form, the raising and lowering operator with respect to index $m$.

\section{Raising and lowering operators for generalized \\classical orthogonal polynomials of continuous \\variable}

Let $y_n (x)$ be an orthogonal polynomials of continuous variable satisfying the differential equation [1]

\begin{equation}
\sigma (x)y''_n (x) + \tau (x)y'_n (x) + \lambda _n y_n (x) = 0,
\end{equation} where $\sigma (x)$ and $\tau (x)$ are polynomials of at most second and first degree, respectively,
and
\begin{equation}
\lambda _n  =  - n\left( {\tau ' + \frac{1} {2}(n - 1)\sigma ''} \right).
\end{equation}

It can be proved that the derivatives of $y_n (x)$, namely, $y_n^{(m)} (x) = v_{mn}(x) ,m = 0,1, \ldots \\n - 1$,
satisfy a similar equation:
\begin{equation}
\sigma (x)v''_{mn} (x) + \tau _m v'_{mn} (x) + \mu _{mn} v_{mn} (x) = 0,
\end{equation} with $\tau _m  = \tau (x) + m\sigma '(x)$ and $\mu _{mn}  =  - (n - m)\left( {\tau ' + \frac{{n + m -
1}} {2}\sigma ''} \right),m = 0,1, \ldots ,n - 1$

We call these polynomials generalized orthogonal polynomials of hypergeometric type, some particular examples of
which are the Legendre and Laguerre, Hermite, Jacobi generalized orthogonal polynomials. [2]

The polynomials of hypergeometric type satisfy an orthognality property with respect to the weight function $\rho
(x)$
\begin{equation}
\int\limits_a^b {y_{\ell}} (x)y_n (x)\rho (x)dx = \delta _{\ell n} d_n^2.
\end{equation}

Similarly the generalized orthogonal polynomials satisfy
\begin{equation}
\int\limits_a^b {v_{m_{\ell}} } (x)v_{mn} (x)\rho _m (x)dx = \delta _{\ell n } d_{mn}^2,
\end{equation} where $d_n^2$ and $d_{mn}^2$ are normalization constants.

It can be proved [3]
\[ d_{mn}^2  = d_{nn}^2 \left( {\prod\limits_{k = m}^{n - 1} {\mu _{kn} } } \right)^{ - 1} ,d_{0n}^2  = d_{nn}^2
\left( {\prod\limits_{k = 0}^{n - 1} {\mu _{kn} } }
\right)^{ - 1},
\] from which $d_{nn}^2$ can be eliminated, therefore
\begin{equation} d_{mn}^2  = d_{0n}^2 \prod\limits_{k = 0}^{m - 1} {\mu _{kn} }.
\end{equation} where $d_{0n}^2$ and $d_n^2$ are given in the tables [6].

The generalized orthogonal polynomials of hypergeometric type can be calculated from the weight function $\rho _m (x)
= \sigma (x)^m \rho (x)$, with the help of the Rodrigues formula:
\begin{equation} v_{mn} (x) = \frac{{A_{mn} B_n }} {{\sigma ^m (x)\rho (x)}}\frac{{d^{n - m} }} {{dx^{n - m}
}}\left\{ {\rho _n (x)} \right\},
\end{equation} where
\begin{equation} A_{mn}  = ( - 1)^m \prod\limits_{k = 0}^{m - 1} {\mu _{kn}  = \frac{{n!}} {{(n - m)!}}}
\prod\limits_{k = 0}^{m - 1} {\left( { - \frac{{\lambda _{n + k} }} {{n + k}}} \right)}. 
\end{equation}

The leading coefficients of the orthogonal polynomial $y_n (x) = a_n x^n  + b_n x^{n - 1}  +  \ldots$
 can be calculated [4]
\begin{equation} a_n  = B_n \prod\limits_{k = 0}^{n - 1} {\left( { - \frac{{\lambda _{n + k} }} {{n + k}}} \right)},
\end{equation} hence, it follows, $A_{nn} B_n  = n!a_n$

We address ourselves to the construction of the raising and lowering operators for the generalized orthogonal
polynomials using the Rodrigues formula as we did in a recent work [5]

We have from (7)
\begin{align}
 \nonumber v_{m,n + 1} (x) &= \frac{{A_{m,n + 1} \;B_{n + 1} }} {{\sigma ^m \rho (x)}}\;\frac{{d^{n + 1 - m} }}
{{dx^{n + 1 - m} }}\left\{ {\rho _n (x)} \right\} = \frac{{A_{m,n + 1} \;B_{n + 1} }} {{\sigma ^m \rho
(x)}}\;\frac{{d^{n - m} }} {{dx^{n - m} }}\left\{ {\tau _n (x)\rho _n (x)} \right\} =  \hfill \\
 \nonumber  &= \frac{{A_{m,n + 1} \;B_{n + 1} }} {{\sigma ^m (x)\rho (x)}}\;\left\{ {\tau _n (x)\frac{{d^{n - m} }}
{{dx^{n - m} }}\left\{ {\rho _n (x)} \right\} + (n - m)\tau '_n \frac{{d^{n - m-1} }} {{dx^{n - m - 1} }}\left\{
{\rho _n (x)} \right\}} \right\} =  \hfill \\
\nonumber   &= \frac{{B_{n + 1} }} {{B_n }}\left\{ {\frac{{A_{m,n + 1} \;}} {{A_{mn} }}\tau _n (x)v_{mn} (x) + (n -
m)\frac{{A_{m,n + 1} \;}} {{A_{m + 1,n} }}\tau '_n \sigma (x)v'_{mn} (x)} \right\} =  \hfill \\
\nonumber   &= \frac{{B_{n + 1} }} {{B_n }}\left\{ {\frac{{n + 1\;}} {{n - m + 1}}\;\frac{n} {{\lambda _n
}}\;\frac{{\lambda _{n + m} }} {{n + m}}\tau _n (x)v_{mn} (x) - \frac{{n + 1\;}} {{n - m + 1}}\;\frac{n} {{\lambda
_n }}\tau '_n \sigma (x)v'_{mn} (x)} \right\}. \hfill \\ 
\end{align} 

The right hand can be considered the raising operator that, when applied to $v_{mn} (x)$
 gives a new polynomial of higher order $v_{m,n + 1} (x)$.

In order to evaluate the lowering operator we need a recurrence relation for the generalized polynomials. We write
\begin{align}
 \nonumber xv_{mn}(x)  &= \sum\limits_{k = 0}^{n + 1} {c_{kn} } v_{mk}(x),\\  c_{kn}  &= \frac{1} {{d_{mk}^2
}}\int\limits_a^b {v_{mk} (x)\;x\;v_{mn} \rho _m (x)dx}. 
\end{align}

From the orthogonality condition (5) we deduce 
\begin{equation*}
 \int\limits_a^b {v_{mn} (x)\;x^r \;\rho _m \;(x)dx = 0} \quad {\rm{for}} \quad r<n-m.
\end{equation*} Since $xp_k^{(m)} (x)$ is a polynomial of order $k+1-m$ it follows that

$c_{kn}  = 0$ if $k+1-m<n-m$, or $k+1<n$. Hence
\begin{equation} xv_{mn}  = \tilde \alpha _n v_{m,n + 1} (x) + \tilde \beta _n v_{mn} (x) + \tilde \gamma _n v_{m,n
+ 1} (x),
\end{equation} where $\tilde \alpha _n  = c_{n + 1,n} ,\quad \tilde \beta _n  = c_{nm} ,\quad   \tilde \gamma _n  =
c_{n - 1,n}$

The coefficients $\tilde \alpha _n ,\;\tilde \beta _n$ and $\tilde \gamma _n$ can be expressed in terms of the
squared norm $d_n^2$ and the leading coefficients $a_n$ and $b_n$ in
$y_n (x)$.

From (11) it can be proved that $d_{mk}^2 c_{kn}  = d_{mn}^2 c_{nk}$.

Since $\tilde \alpha _{n - 1}  = c_{n,n - 1} ,\;\tilde \gamma _{n}  = c_{n - 1,n}$, if we put $k=n-1$ we obtain
\begin{equation*} c_{n - 1,n} d_{m,n - 1}^2  = c_{n,n - 1} d_{mn}^2, 
\end{equation*} hence 
\begin{equation*}
\tilde \gamma _n  = \tilde \alpha _{n - 1} \;\frac{{d_{m,n}^2 }} {{d_{m,n - 1}^2 }}.
\end{equation*}

Introducing the expansion $y_n (x) = a_n x^n  + b_n x^{n - 1}  +  \ldots$ in (12) and comparing the coefficients of
the highest terms, we have
\begin{align*} &a_n (n - m + 1) = \tilde \alpha _n a_{n + 1} (n + 1), \\[1ex] &b_n (n - m) = \tilde \alpha _n b_{n +
1} n + \tilde \beta _n a_n n.
\end{align*}

Hence
\begin{align}
\tilde \alpha _n  &= \frac{{a_n }} {{a_{n + 1} }}\;\frac{{n - m + 1}} {{n + 1}},\\
\tilde \beta _n  &= \frac{{b_n }} {{a_n }}\;\frac{{(n - m)}}
  {n} - \frac{{b_{n  + 1}}} {{a_{n  + 1}}}\;\frac{{n + 1 - m}} {{n + 1}},\\
\tilde \gamma _n  &= \frac{{a_{n - 1} }} {{a_n }}\;\frac{{n - m}} {n}\frac{{d_{m,n}^2 }} {{d_{m,n - 1}^2 }}.
\end{align}

Substituting (9) in $\tilde \alpha _n$ we obtain
\begin{equation}
\tilde \alpha _n  = \frac{{ - B_n }} {{B_{n + 1} }}\;\frac{{n - m + 1}} {{n + 1}}\;\frac{{\lambda _n }}
{n}\;\frac{{2n}} {{\lambda _{2n} }}\;\frac{{2n + 1}} {{\lambda _{2n + 1} }}.
\end{equation}

Hence (10) can be written
\begin{equation}
\tilde \alpha _n \;\frac{{\lambda _{2n} }} {{2n}}\;v_{m,n + 1} (x) = \left\{ {\frac{{\lambda _{n + m} }} {{n +
m}}\;\frac{{\tau _n (x)}} {{\tau '_n }}\;v_{mn} (x) - \sigma (x)\;v'_{mn} (x)} \right\}.
\end{equation}

Inserting (12) in (17) we obtain
\begin{equation}
\tilde \gamma _n \;\frac{{\lambda _{2n} }} {{2n}}\;v_{m,n - 1} (x) = \left\{ - \frac{{\lambda _{n + m} }} {{n +
m}}\;\frac{{\tau _n (x)}} {{\tau '_n }}+ \frac{{\lambda _{2n} }} {{2n}}\left( {x - \tilde \beta _n } \right)\right\}
v_{mn} (x) + \sigma (x)\;v'_{mn} (x)
\end{equation}

The right hand side of (17) and (18) can be considered the raising and lowering operators for the generalized
classical orthogonal polynomials with respect to the index $n$.

All the constants $\tilde \alpha _n ,\tilde \beta _n ,\tilde \gamma _n ,\lambda _n ,\tau '_n$ can be calculated from
the tables [6].

Now we define the orthonormalized function
\begin{equation}
\psi _{mn}(x)=d_{mn}^{-1}\sqrt {\rho _m(x)}v_{mn}(x),
\end{equation} hence
\begin{equation}
\psi '_{mn}(x)={1 \over 2}{{\rho '_m(x)} \over {\rho (x)}}\psi _{mn}(x)+d_{mn}^{-1}\sqrt {\rho _m(x)}v'_{mn}(x)={1
\over 2}{{\tau _{m-1}(x)} \over {\sigma (x)}}\psi _{mn}(x)+d_{mn}^{-1}\sqrt {\rho _m(x)}v'_{mn}(x)
\end{equation}

Multiplying (17) by $d_{mn}^{-1}\sqrt {\rho _m(x)}$ and substituting (20) in (17) we get
\begin{align}
\nonumber \tilde \alpha _n{{\lambda _{2n}} \over {2n}}{{d_{m,n+1}} \over {d_{mn}}}\psi _{m,n+1}(x) &={{\lambda
_{n+m}}
\over {n+m}}{{\tau _n(x)} \over {\tau '_n}}\psi _{mn}(x)+{1 \over 2}\tau _{m-1}(x)\psi _{mn}(x)-\sigma (x)\psi
'_{mn}(x)=\\
&=L^+(x,n)\psi_{m,n}(x)
\end{align}

Similarly 
\begin{align}
\nonumber\tilde \gamma _n{{\lambda _{2n}} \over {2n}}{{d_{m,n-1}} \over {d_{mn}}}\psi _{m,n-1}(x)= &
\left\{ {-{{\lambda _{n+m}} \over {n+m}}{{\tau _n(x)} \over {\tau '_n}}+{{\lambda _{2n}} \over {2n}}\left( {x-\tilde
\beta _n} \right)-{1 \over 2}\tau _{m-1}(x)} \right\}\psi _{mn}(x)+ \\
 & +\sigma (x)\psi '_{mn}(x)=L^-(x,n)\psi_{m,n}(x)
\end{align} that can be considered the raising and lowering operators for the generalized orthonormal functions
$\psi_{mn}(x)$. These operators are mutually adjoint with respect to the scalar product of unit weight.

Following the same procedure as in [5] we can factorize the raising and lowering operators as follows:
\begin{equation*}
L^-(x,n+1)L^+(x,n)=\mu (n)-\sigma (x)H(x,n)
\end{equation*}
\begin{equation*}
L^+(x,n)L^-(x,n+1)=\mu (n)-\sigma (x)H(x,n+1)
\end{equation*}
where
$$\mu (n)={{\lambda _{2n}} \over {2n}}\,{{\lambda _{2n+2}} \over {2n+2}}\,\tilde \alpha _n\,\tilde \gamma _{n+1}$$
and $H(x,n)$ is the differential operator derived from the left hand side of (3) after substituting $\psi_{mn}(x)$
instead of $v_{mn}(x)$.

Notice that the factorization of the raising and lowering operators is defined in a basis independent manner, which
is equivalente to the Infeld-Hull method.

\section{Raising and lowering operators for generalized \\classical orthogonal polynomials of discrete \\variable on
uniform lattice}

Let $y_n (x)$ be an orthogonal polynomial of discrete variable satisfying the difference equation [7]
\begin{equation}
\sigma (x)\;\Delta \nabla \;y_n (x) + \tau (x)\;\Delta y_n (x) + \lambda _n y_n (x) = 0,\;
\end{equation} where $\sigma (x)$ and $\tau (x)$ are polynomials of at most of second and first degree, respectively,
\begin{equation}
\lambda _n  =  - n\left( {\tau ' + \frac{1} {2}(n - 1)\sigma ''} \right),
\end{equation} and the forward and backward difference operators are, respectively,
\begin{equation*}
\Delta f(x) = f(x + 1) - f(x),\quad \nabla f(x) = f(x) - f(x - 1).
\end{equation*}

It can be proved [8] that the differences of $y_n (x)$, namely $\Delta ^m y_n (x)=v_{mn}(x)$ satisfy similar
equation of hypergeometric type:
\begin{equation}
\sigma (x)\;\Delta \nabla \;v_{mn} (x) + \tau _m (x)\;\Delta v_{mn} (x) + \mu _{mn} v_{mn} (x) = 0\;
\end{equation} with 
\begin{align*}
\tau _m (x) &= \tau (x + m) + \sigma (x + m) - \sigma (x),\\
\mu _{mn} & = \lambda _n  - \lambda _m  =  - (n - m)\left( {\tau ' + \frac{{n + m - 1}} {2}\sigma ''} \right),\quad
m = 0,1, \ldots n - 1.
\end{align*}

We call the polynomials $v_{mn}(x)$ the generalized classical orthogonal polynomials of discrete variable, among
them we find the Hahn, Chebyshev, Meixner, Kravchuk and Charlier polynomials.

The classical orthogonal polynomials of discrete variable satisfy an orthogonality property with respect to the
weight function $\rho (x)$
\begin{equation}
\sum\limits_{x = a}^{b - 1} {y_{\ell} } (x)y_n (x)\rho (x) = \delta _{\ell n } d_n^2. 
\end{equation}

Similarly the generalized classical orthogonal polynomials of discrete variable satisfy the orthogonality property
\begin{equation}
\sum\limits_{x = a}^{b - 1} {v_{m \ell}} (x)v_{mn} (x)\rho _m (x) = \delta _{\ell n } d_{mn}^2, 
\end{equation} where $d_n^2$ and $d_{mn}^2$ are normalization constants. It can be proved [9]
\begin{equation*} d_{mn}^2  = d_{nn}^2 \left( {\prod\limits_{k = m}^{n - 1} {\mu _{kn} } } \right)^{ - 1} \quad
,\quad d_{0n}^2 =d_{nn}^2 \left( {\prod\limits_{k = 0}^{n - 1} {\mu _{kn} } }
\right)^{ - 1}. 
\end{equation*}

If we eliminate $d_{nn}$ in the above equations we get 
\begin{equation} d_{mn}^2  = d_{0n}^2 \prod\limits_{k = 0}^{m - 1} {\mu _{kn} }.
\end{equation}

The generalized clasical orthogonal polynomials of discrete variable can be calculated from the weight function $\rho
_m (x)$ with the formula [9]:
\begin{equation} v_{mn} (x) = \frac{{A_{mn} B_n }} {{\rho _m (x)}}\nabla ^{n - m} \left\{ {\rho _n (x)} \right\},
\end{equation} where
\begin{align} A_{mn} & = \frac{{n!}} {{(n - m)!}}\prod\limits_{k = 0}^{m - 1} {\left( { - \frac{{\lambda _{n + k} }}
{{n + k}}} \right)}, \\ B_n  &= \frac{{\Delta ^n y_n (x)}} {{A_{nn} }}.
\end{align}

The leading coefficients of the classical orthogonal polynomial of discrete variable $y_n (x) = a_n x^n  + b_n x^{n
- 1}  +  \ldots$, are given by [10]
\begin{equation} a_n  = B_n \prod\limits_{k = 0}^{n - 1} {\left( { - \frac{{\lambda _{n + k} }} {{n + k}}} \right)} 
\end{equation} from which it follows $A_{nn} B_n  = n!a_n$.

We have now all the necessary ingredients to construct the raising and lowering operators for the generalized
orthogonal polynomials of discrete variable in analogy with those of continuous variable. From (29) we have

\begin{align}
\nonumber v_{m,n+1}(x)&={{A_{m,n+1}\;B_{n+1}} \over {\rho _m(x)}}\nabla ^{n-m+1}\left\{ {\rho _{n+1}(x)} \right\}=\\
\nonumber&={{A_{m,n+1}\;B_{n+1}} \over {\rho _m(x)}}\nabla ^{n-m}\left\{ {\Delta \rho _{n+1}(x-1)} \right\}=\\
\nonumber&={{A_{m,n+1}\;B_{n+1}} \over {\rho _m(x)}}\nabla ^{n-m}\left\{ {\tau _n(x)\rho _n(x)} \right\}=\\
&={{A_{m,n+1}\;B_{n+1}} \over {\rho _m(x)}}\left\{ {\tau _n(x)\nabla ^{n-m}\rho _n(x)+(n-m)\tau '_n\nabla
^{n-m-1}\rho _n(x-1)} \right\}
\end{align}

From (29) we have
\begin{align*}
\nabla ^{n - m} \left\{ {\rho _n (x)} \right\} &= \frac{{\rho _m (x)}} {{A_{mn} \;B_n }}v_{mn}(x) \\
\nabla ^{n - m - 1} \left\{ {\rho _n (x - 1)} \right\} &= \frac{{\sigma (x)\rho _m (x)}} {{A_{m + 1,n} \;B_n
}}\Delta ^{m + 1} y_n (x - 1) = \frac{{\sigma (x)\rho _m (x)}} {{A_{m + 1,n} \;B_n }} = \nabla {v_{mn} (x)}
\end{align*}

Substituting the last two expressions in (33) and using (30) we obtain
\begin{equation} v_{m,n + 1} (x) = \frac{{B_{n + 1} }} {{B_n }}\left\{ {\frac{{n + 1}} {{n + 1 - m}}\;\frac{{\lambda
_{n + m} }} {{n + m}}\;\frac{n} {{\lambda _n }}\tau _n (x)v_{mn} (x) - \frac{{n + 1}} {{n + 1 - m}}\;\frac{n}
{{\lambda _n }}\tau '_n \sigma (x)\nabla v_{mn} (x)} \right\}
\end{equation} that raises in one step the order of the generalized polynomials in terms of the polynomials $v_{mn}
(x)$ and $\nabla v_{mn} (x)$

In order to evaluate the lowering operator we calculate a recurrence relation for the generalized orthogonal
polynomials of discrete variable. We write
\begin{align}
\nonumber x\;v_{mn} (x) &= \sum\limits_{k = 0}^{n + 1} {c_{kn} v_{mk} (x)}, \\ c_{kn}  &= \frac{1} {{d_{mk}^2
}}\sum\limits_{x = a}^{b - 1} {v_{mk} (x)} \;x\;v_{mn} (x)\rho _m (x).
\end{align}

As in the case of the continuous variable $c_{kn}  = 0$, if $k+1<n$. Hence
\begin{equation} x\;v_{mn} (x) = \tilde \alpha _n v_{m,n + 1} (x) + \tilde \beta _n v_{mn} (x) + \tilde \gamma _n
v_{m,n - 1} (x)
\end{equation} where $\tilde \alpha _n  = c_{n + 1,n} ,\quad \tilde \beta _n  = c_{nn} ,\quad \tilde \gamma _n  =
c_{n - 1,n}$

From (35) it follows that $d_{mk}^2 c_{kn}  = d_{mn}^2 c_{nk}$

Since $\tilde \alpha _{n - 1}  = c_{n,n - 1} ,\;\tilde \gamma _n  = c_{n - 1,n}$, if we put $k=n-1$, we get $c_{n -
1,n} d_{m,n - 1}^2  = c_{n,n - 1} d_{m,n}^2$, hence
\begin{equation*}
\tilde \gamma _n  = \tilde \alpha _{n - 1} \;\frac{{d_{m,n}^2 }} {{d_{m,n - 1}^2 }}.
\end{equation*}

Introducing the expansion $y_n (x) = a_n x^n  + b_n x^{n - 1}  +  \ldots$ in (36), comparing the coefficients of the
highest terms, and using
$$\Delta ^mx^n=n(n-1)\ldots (n-m+1)x^{n-m}+{m \over 2}n(n-1)\ldots (n-m)x^{n-m+1}+\ldots $$

we obtain
\begin{align*} &a_n (n - m + 1) = \tilde \alpha _n a_{n + 1} (n + 1)\\ &a_nn(n-m){m \over 2}+b_n(n-m)=\tilde \alpha
_nb_{n+1}n+\tilde \beta _na_nn+\tilde \alpha _na_{n+1}(n+1)n{m \over 2}.
\end{align*}

From these relations and (32) we obtain
\begin{align}
\tilde \alpha _n  &= \frac{{a_n }} {{a_{n + 1} }}\;\frac{{n - m + 1}} {{n + 1}} = -\frac{{B_n }} {{B_{n + 1}
}}\;\frac{{\lambda _n }} {n}\frac{{2n}} {{\lambda _{2n} }}\frac{{2n + 1}} {{\lambda _{2n + 1} }}\frac{{n - m + 1}}
{{n + 1}},\\
\tilde \beta _n  &= \frac{{b_n }} {{a_n }}\;\frac{{(n - m)}} {n} - \frac{{b_{n + 1} }} {{a_{n + 1} }}\;\frac{{n - m
+ 1}} {{n + 1}} - \frac{m} {2}, \\
\tilde \gamma _n  &= \frac{{a_{n - 1} }} {{a_n }}\;\frac{{n - m}} {n}\frac{{d_{m,n}^2 }} {{d_{m,n - 1}^2 }}.
\end{align}

Hence (34) can be written
\begin{equation}
\tilde \alpha _n\frac{{\lambda _{2n} }} {{2n}}\;v_{m,n + 1} (x) = \frac{{\lambda _{n + m} }} {{n + m}}\frac{{\tau _n
(x)}} {{\tau '_n }}v_{mn} (x) - \sigma (x)\nabla v_{mn} (x)
\end{equation}

Inserting (36) in (40) we get
\begin{equation}
\tilde \gamma _n \;\frac{{\lambda _{2n} }} {{2n}}\;v_{m,n - 1} (x) =  - \frac{{\lambda _{n + m} }} {{n +
m}}\;\frac{{\tau _n (x)}} {{\tau '_n }}\;v_{mn} (x) + \frac{{\lambda _{2n} }} {{2n}}\left( {x - \tilde \beta _n }
\right)v_{mn} (x) + \sigma (x)\;\nabla v_{mn} (x)
\end{equation}

The right side of (40) and (41) can be considered the raising and lowering operators with respect to the index $n$
for the generalized orthogonal polynomials of discrete variable on homogeneous lattice.

All the constants $\tilde \alpha _n ,\tilde \beta _n ,\tilde \gamma _n ,\lambda _n ,\tau '_n$ can be calculated from
the tables [11]. Obviously, when $m=0$, $\tilde \alpha _n ,\tilde \beta _n ,\tilde \gamma$ become, respectively,
$\alpha _n,\beta _n ,\gamma_n$.

Now we define the orthonormal function of discrete variable 
\begin{equation*}
\phi _{mn}(x)=d_{mn}^{-1}\sqrt {\rho _m(x)}v_{mn}(x)
\end{equation*}

Using the identity $\displaystyle {{\nabla \rho _m(x)} \over {\rho _m(x)}}={{\tau _{m-1}(x)} \over {\tau
_{m-1}(x)+\sigma (x)}}$ and the properties of the backwards operator we get
\begin{align}
\nonumber \nabla \phi _{mn}(x)&=\sqrt {{{\sigma (x)} \over {\tau _{m-1}(x)+\sigma (x)}}}\;d_{mn}^{-1}\sqrt {\rho
_m(x)}\;\nabla v_{mn}(x)+\\ &+{{\tau _{m-1}(x)} \over {\sqrt {\sigma (x)}+\sqrt {\tau _{m-1}(x)+\sigma
(x)}}}\;{{\phi _{mn}(x)} \over {\sqrt {\sigma (x)+\tau _{m-1}(x)}}}
\end{align}

Multiplying both sides of (40) by $d_{mn}^{-1}\sqrt {\rho _m(x)}$ and inserting the value  $d_{mn}^{-1}\sqrt {\rho
_m(x)}$ $\nabla v_{mn}(x)$ obtained in (42), we get
\begin{align}
\nonumber &\tilde \alpha _n\;{{\lambda _{2n}} \over {2n}}\;{{d_{m,n+1}} \over {d_{m,n}}}\phi _{m,n+1}(x)=
L^+(x,n)\phi _{mn}(x)=\\
\nonumber &=\left\{ {+{{\lambda _{n+m}} \over {n+m}}\;{{\tau _n(x)} \over {\tau '_n}}+{{\sqrt {\sigma (x)}\;\tau
_{m-1}(x)} \over {\sqrt {\sigma (x)}+\sqrt {\sigma (x)+\tau _{m-1}(x)}}}}
\right\}\phi _{mn}(x)+\\ &-\sqrt {\sigma (x)-\left( {\tau _{m-1}(x)+\sigma (x)} \right)}\;\nabla \phi _{mn}(x)
\end{align}

Similarly
\begin{align}
\nonumber & \tilde \gamma _n\;{{\lambda _{2n}} \over {2n}}\;{{d_{m,n-1}} \over {d_{m,n}}}\phi _{m,n-1}(x)=
L^-(x,n)\phi _{mn}(x)=\\
\nonumber & =\left\{ {-{{\lambda _{n+m}} \over {n+m}}\;{{\tau _n(x)} \over {\tau '_n}}+{{\lambda _{2n}} \over
{2n}}\left( {x-\tilde \beta _n} \right)-{{\sqrt {\sigma (x)}\;\tau _{m-1}(x)} \over {\sqrt {\sigma (x)}+\sqrt
{\sigma (x)+\tau _{m-1}(x)}}}} \right\}\phi _{mn}(x)+\\ &\sqrt {\sigma (x)\;\left( {\tau _{m-1}(x)+\sigma (x)}
\right)}\;\nabla \phi _{mn}(x)
\end{align}

The expressions (43) and (44) can be considered the raising and lowering operators, respectively, for the generalized
orthonormal functions on homogeneous lattice. These operators are mutually adjoint with respect to the scalar
product of unit weight.

Notice that in (43) and (44) the last term is proportional to $\nabla \phi _{mn}(s)$, which in the continuous limit
becomes the derivative $\psi '_{mn}(x)$.

As in [5] we can factorize the raising and lowering operators as follows:
$$L^-(x,n+1)L^+(x,n)=\mu (n)+\mu (x+1,n)H(x,n)$$
$$L^+(x,n)L^-(x,n+1)=\mu (n)+\mu (x,n-1)H(x,n+1)$$
where 
$$\mu (n)={{\lambda _{2n}} \over {2n}}\,{{\lambda _{2n+2}} \over {2n+2}}\,\tilde \alpha _n\,\tilde \gamma
_{n+1},$$
$$\mu (x,n)={{\lambda _n} \over n}\,{{\tau _n(x)} \over {\tau '_n}}-\sigma (x)$$
and $H(x,n)$ is the difference operator derived from the left hand side of (25) after substituting $\phi _{mn}(x)$
instead of $v_{mn}(x)$.

\section{Raising and lowering operators for classical ortho\-go\-nal polynomials of a discrete variable on
nonuniform lattice}

Let $y (s)$ a function of discrete variable satisfying the difference equation with respect to the lattice function
$x(s)$
\begin{equation}
\sigma (s)\frac{\Delta } {{\Delta x\left( {s - {1 \mathord{\left/
 {\vphantom {1 2}} \right.
 \kern-\nulldelimiterspace} 2}} \right)}}\left\{ {\frac{{\nabla y(s)}} {{\Delta x(s)}}} \right\} + \tau
(s)\frac{{\Delta y(s)}} {{\Delta x(s)}} + \lambda y(s) = 0
\end{equation} where $\sigma (s) \equiv \sigma \left[ {x(s)} \right]\;,\quad \tau (s) \equiv \tau \left[ {x(s)}
\right]$ are functions of $x(s)$ of at most of second and first degree, respectively.

It can be proved [12] that the functions $v_k (s)$ connected with the solutions $y(s)$ by the relations
\begin{align} v_k (s) &= \frac{{\Delta v_{k - 1} (s)}} {{\Delta x_{k - 1} (s)}} ,\quad v_0 (s) = y(s) \\
\nonumber x_k (s) &= x\left( {s + \frac{k}{2}} \right),\quad k = 0,1,2 \ldots 
\end{align} satisfy the difference equation
\begin{equation}
\sigma (s)\frac{\Delta } {{\Delta x_k (s - {1 \mathord{\left/
 {\vphantom {1 2}} \right.
 \kern-\nulldelimiterspace} 2})}}\left\{ {\frac{{\nabla v_k (s)}} {{\nabla x_k (s)}}} \right\} + \tau _k
(s)\frac{{\Delta v_k (s)}} {{\Delta x_k (s)}} + \mu _k v_k (s) = 0
\end{equation} where 
\begin{align}
\tau _k (s) &= \frac{{\sigma (s + k) - \sigma (s) + \tau (s + k)\Delta x\left( {s + k - {1 \mathord{\left/
 {\vphantom {1 2}} \right.
 \kern-\nulldelimiterspace} 2}} \right)}} {{\Delta x\left( {{{s + (k - 1)} \mathord{\left/
 {\vphantom {{s + (k - 1)} 2}} \right.
 \kern-\nulldelimiterspace} 2}} \right)}} \\
\mu _k  &= \lambda  + \sum\limits_{m = 0}^{k - 1} {\frac{{\Delta \tau _m (s)}} {{\Delta x_m (s)}}}  = \lambda  +
\sum\limits_{m = 0}^{k - 1} {\tau '_m } 
\end{align} provided the lattice functions $x(s)$ have the form
\begin{align} x(s)&=c_1s^2+c_2s+c_3 \quad {\rm or} \\
 x(s)&=c_1q^s+c_2q^{-s}+c_3  
\end{align} with $c_1, c_2, c_3, q$, arbitrary constants.

When $\mu _k=0$ for $k=n$ in (47) $v_n={\rm const}$. It can be proved that when $k<n$, \, $v_k(s)$ is a polynomial
in $x_k(s)$ and in particular for $k=0$, $v_0(s)=y(s)$ is a polynomial of degree $n$ in $x(s)$ satisfying (45).

An explicit expression for $\lambda_n$, when $\mu_n=0$, is given by 
\begin{equation}
\lambda _n=-{{{\rm sh}n\omega } \over {{\rm sh}\omega }}\left\{ {{\rm ch}(n-1)\omega\;  \tau '+{1 \over 2}{{{\rm
sh}(n-1)\omega } \over {{\rm sh}\omega }}\sigma ''} \right\}
\end{equation} where $\omega ={1 \over 2}{\rm ln}q$, or $q=e^{2\omega }$. For the square lattice (50) $\omega=0$;
and for the $q-$lattice (51) we have
\begin{equation*} {{{\rm sh}n\omega } \over {{\rm sh}\omega }}={{q^{{n \mathord{\left/ {\vphantom {n 2}} \right.
\kern-\nulldelimiterspace} 2}}-q^{{{-n} \mathord{\left/ {\vphantom {{-n} 2}} \right.
\kern-\nulldelimiterspace} 2}}} \over {q^{{1 \mathord{\left/ {\vphantom {1 2}} \right. \kern-\nulldelimiterspace}
2}}-q^{{{-1} \mathord{\left/ {\vphantom {{-1} 2}} \right.
\kern-\nulldelimiterspace} 2}}}}\equiv \left[ n \right]_q
\end{equation*} The polynomials solutions of (45) satisfy the following orthogonality condition with respect to the
weight functions $\rho (s)$, namely,
\begin{equation}
\sum\limits_{s=a}^{b-1} {y_\ell (s)y_n(s)\rho (s)\Delta x\left( {s-{1 \over 2}} \right)=\delta _{\ell n}d_n^2}
\end{equation}

Similarly for the differences of the polynomials $y_n(s)$, namely,
\begin{equation*} v_{mn}(s)\equiv \Delta ^{(m)}\left[ {y_n(s)} \right]=\Delta _{m-1}\;\Delta _{m-2}\ldots \Delta
_0\left[ {y_n(s)} \right],\quad \Delta _k\equiv {\Delta  \over {\Delta x_k(s)}}
\end{equation*} it holds
\begin{equation}
\sum\limits_{s=a}^{b-k-1} {v_{m\ell }(s)v_{mn}(s)\rho _m(s)\Delta x_m\left( {s-{1 \over 2}} \right)=\delta _{\ell
n}d_{mn}^2}
\end{equation} where $\rho _m(s)=\rho (s+m)\prod\limits_{i=1}^m {\sigma (s+i)}$

It can be proved that the normalization constants satisfy
\begin{equation*} d_{mn}^2=d_{nn}^2\left( {\prod\limits_{k=m}^{n-1} {\mu _{kn}}} \right)^{-1},\quad
d_{0n}^2=d_{nn}^2\left( {\prod\limits_{k=0}^{n-1} {\mu _{kn}}} \right)^{-1}
\end{equation*} from which $d_{nn}^2$ can be eliminated:
\begin{equation} d_{mn}^2=d_{0n}^2\prod\limits_{k=o}^{m-1} {\mu _{kn}}
\end{equation}

A particular solution of (45) when $\lambda=\lambda_n$ is given by the Rodrigues type formula
\begin{equation} {y_n(s)} ={{B_n} \over {\rho (s)}}\nabla _n^{(n)}\left[ {\rho _n(s)} \right]={{B_n} \over {\rho
(s)}}{\nabla  \over {\nabla x_1(s)}}\cdots {\nabla  \over {\nabla x_n(s)}}\left[ {\rho _n(s)} \right]
\end{equation}

A solution of (47) when $\mu_k$ is restricted to $\lambda_n$, namely, $\mu _{mn}=\mu _m(\lambda _n)=\lambda
_n-\lambda _m,$ $ 0,1,\ldots n-1$, is given by:
\begin{equation} v_{mn}(s)={{A_{mn}B_n} \over {\rho _m(s)}}\nabla _n^{(n-m)}\left[ {\rho _n(s)} \right]={{A_{mn}B_n}
\over {\rho_m (s)}}{\nabla  \over {\nabla x_{m+1}(s)}}\cdots {\nabla  \over {\nabla x_{n-1}(s)}}{\nabla  \over
{\nabla x_n(s)}}\left[ {\rho _n(s)} \right]
\end{equation} where
\begin{align} A_{mn}&=(-1)^m\prod\limits_{k=o}^{m-1} {\mu _{kn}}={{\left[ n \right]!} \over {\left[ {n-m}
\right]!}}\prod\limits_{k=o}^{m-1} {{{\lambda _{n+k}} \over {\left[ {n+k} \right]}}}\\
\nonumber B_n&=A_{nn}^{-1}\Delta ^{(n)}y_n(s)
\end{align}

Formulas (56) and (57) can be written in terms of the mean difference operator [13]
$\delta f(s)=f\left( {s+{1 \over 2}} \right)-f\left( {s-{1 \over 2}} \right)=\Delta f\left( {s-{1 \over 2}}
\right)=\nabla f\left( {s+{1 \over 2}} \right)$, that is to say,
\begin{align} y_n(s)&={{B_n} \over {\rho (s)}}\left[ {{\delta  \over {\delta x(s)}}} \right]^n\rho_n\left( {s-{n
\over 2}} \right)\\ v_{mn}(s)&={{A_{mn}B_n} \over {\rho _m(s)}}\left[ {{\delta  \over {\delta x\left( {s+{m \over
2}} \right)}}} \right]^{n-m}\rho_n\left( {s-{n \over 2}+{m \over 2}} \right)
\end{align}

In order to obtain the raising and lowering operators of the classical orthogonal polynomials on non-homogeneous
lattice, we apply the Rodrigues formula (56)
\begin{equation*} y_{n+1}(s)={{B_{n+1}} \over {\rho (s)}}\nabla _{n+1}^{(n+1)}\left\{ {\rho _{n+1}(s)}
\right\}={{B_{n+1}} \over {\rho (s)}}{\nabla  \over {\nabla x_1(s)}}\cdots {\nabla  \over {\nabla
x_{n+1}(s)}}\left\{ {\rho _{n+1}(s)} \right\}
\end{equation*}

Since
\begin{equation*} {{\nabla \rho _{n+1}(s)} \over {\nabla x_{n+1}(s)}}={{\Delta \rho _{n+1}(s-1)} \over {\Delta
x_{n+1}(s-1)}}={{\Delta \left\{ {\sigma (s)\rho _n(s)} \right\}} \over {\Delta x_n\left( {s-{1 \over 2}}
\right)}}=\tau _n(s)\rho _n(s),
\end{equation*} using (59) we have
\begin{align}
\nonumber y_{n+1}(s)={{B_{n+1}} \over {\rho (s)}}\nabla _n^{(n)}\left\{ {\tau _n(s)\rho _n(s)} \right\}={{B_{n+1}}
\over {\rho (s)}}\left[ {{\delta  \over {\delta x(s)}}}
\right]^n\left\{ {\tau _n\left( {s-{n \over 2}} \right)\rho _n\left( {s-{n \over 2}} \right)} \right\}= \\
={{B_{n+1}} \over {\rho (s)}}\left\{ {\tau _n(s)\left[ {{\delta  \over {\delta x(s)}}} \right]^n\rho _n\left( {s-{n
\over 2}} \right)+{{{\rm sh}n\omega } \over {{\rm sh}\omega }}\tau '_n\left[ {{\delta  \over {\delta x\left( {s-{1
\over 2}} \right)}}} \right]^{n-1}\rho _n\left( {s-{n
\over 2}-{1
\over 2}} \right)} \right\}
\end{align}

The last step can be proved by induction for both cases of $x(s)$ on non-homogeneous lattice (50) and (51).

First of all, we transform the properties of these functions [14] given by
\begin{align*} x(s+n)-x(s)&={{{\rm sh}n\omega } \over {{\rm sh}\omega }}\nabla x\left( {s+{{n+1} \over 2}} \right)\\ 
x(s+n)+x(s)&={\rm ch}n\omega \;x(s)+ {\rm const}.
\end{align*} into the difference relations
\begin{align} &\delta x\left( {s+{n \over 2}} \right)-\delta x\left( {s-{n \over 2}} \right)={{{\rm sh}n\omega }
\over {{\rm sh}\omega }}\left\{ {\delta x\left( {s+{1 \over 2}} \right)-\delta x\left( {s-{1 \over 2}} \right)}
\right\}\\ &{1 \over 2}\left\{ {\delta x\left( {s+{n \over 2}} \right)+\delta x\left( {s-{n \over 2}} \right)}
\right\}={\rm ch}n\omega \;\delta x(s)
\end{align}

Suppose it is true that for any two functions of discrete variable it holds
\begin{align*}
\left( {{\delta  \over {\delta x(s)}}} \right)^n \left\{ f(s)g(s) \right\} &=f\left( {s+{n \over 2}} \right)\left(
{{\delta  \over {\delta x(s)}}} \right)^ng(s)+\\ &+{{{\rm sh}n\omega } \over {{\rm sh}\omega }}{{\delta f\left(
{s+{{n-1} \over 2}} \right)} \over {\delta x\left( {s+{{n-1} \over 2}} \right)}}\left( {{\delta  \over {\delta
x\left( {s-{1 \over 2}} \right)}}} \right)^{n-1}g\left( {s-{1
\over 2}} \right)+\ldots 
\end{align*}

Then using the properties of the mean operator we have:
\begin{align*}
\left( {{\delta  \over {\delta x(s)}}} \right)^{n+1}\left\{ f(s)g(s)\right\} &=f\left( {s+{{n+1} \over 2}}
\right)\left( {{\delta  \over {\delta x(s)}}} \right)^{n+1}g(s)+\\ &+{{\delta f\left( {s+{n \over 2}}
\right)} \over {\delta x(s)}}\left( {{\delta  \over {\delta x\left( {s-{1 \over 2}} \right)}}} \right)^ng\left(
{s-{1 \over 2}} \right)+\\ &+{{{\rm sh}n\omega } \over {{\rm sh}\omega }}{{\delta f\left( {s+{n \over 2}} \right)}
\over {\delta x\left( {s+{n \over 2}} \right)}}{\delta  \over {\delta x(s)}}\left( {{\delta  \over {\delta x\left(
{s-{1 \over 2}} \right)}}}
\right)^{n-1}g\left( {s-{1 \over 2}} \right)+\ldots 
\end{align*}

The second and third term on the right hand side can we written:
\begin{equation*} {{\delta f\left( {s+{n \over 2}} \right)} \over {\delta x\left( {s+{n \over 2}} \right)}}\left\{
{{{\delta x\left( {s+{n \over 2}} \right)} \over {\delta x(s)}}+{{{\rm sh}n\omega } \over {{\rm sh}\omega }}{{\delta
x\left( {s-{1 \over n}} \right)} \over {\delta x(s)}}} \right\}\left( {{\delta  \over {\delta x\left( {s-{1 \over
2}} \right)}}} \right)^ng\left( {s-{1 \over 2}}
\right)
\end{equation*}

Using (62) and (63) the expression between curly brackets is equal to ${\rm sh}(n+1)\omega/{\rm sh}\omega$, therefore
\begin{align*}
\left( {{\delta  \over {\delta x(s)}}} \right)^{n+1}\left\{ f(s)g(s)\right\} &=f\left( {s+{{n+1} \over 2}}
\right)\left( {{\delta  \over {\delta x(s)}}} \right)^{n+1}g(s)+\\ &+{{{\rm sh}(n+1)\omega } \over {{\rm sh}\omega
}}{{\delta f\left( {s+{n \over 2}} \right)} \over {\delta x\left( {s+{n \over 2}} \right)}}\left( {{\delta  \over
{\delta x\left( {s-{1 \over 2}} \right)}}} \right)^ng\left( {s-{1 \over 2}} \right)+\ldots 
\end{align*} as required. Substituting $f(x)=\tau _n\left( {s-{n \over 2}} \right)$ and $g(s)=\rho _n\left( {s-{n
\over 2}} \right)$, the terms of lower degree become zero, due to the properties of function $\tau_n(s)$. Therefore
(61) is proved.

Using (60) for $m=1$ we have
\begin{equation*} {{\nabla y_n(s)} \over {\nabla x(s)}}={{\Delta y_n(s-1)} \over {\Delta
x(s-1)}}=v_{1n}(s-1)={{A_{1n}B_n} \over {\rho _1(s-1)}}\left( {{\delta  \over {\delta x\left( {s-{1 \over 2}}
\right)}}} \right)^{n-1}\rho _n\left( {s-{n \over 2}-{1 \over 2}} \right)
\end{equation*}

Therefore, (61) can be written
\begin{equation} y_{n+1}(s)={{B_{n+1}} \over {B_n}}\left\{ {\tau _n(s)y_n(s)+{{{\rm sh}n\omega } \over {{\rm
sh}\omega }}{{\tau '_n} \over {A_{1n}}}\sigma (s){{\nabla y_n(s)} \over {\nabla x(s)}}} \right\}
\end{equation}

Alvarez-Nodarse and Costas-Santos [15] [16] have given the same formula for the lattice (51). Here we have proved
the similar expression for both cases (50) and (51).

From (64) we can calculate the raising and lowering operators. Instead, we proceed to the general case in section 5,
and then take the value $m=0$.

\section{Raising and lowering operators for generalized cla-ssical orthogonal polynomials of discrete variable on
non-uniform lattice}

From (57) and (60) we obtain
\begin{align*} v_{m,n+1}(s)&={{A_{m,n+1}\;B_{n+1}} \over {\rho _m(s)}}\nabla _{n+1}^{(n+1-m)}\left\{ {\rho
_{n+1}(s)} \right\}= \\ &={{A_{m,n+1}\;B_{n+1}} \over {\rho _m(s)}}\nabla _n^{(n-m)}\left\{ {\tau _n(s)\rho _n(s)}
\right\}= \\ &={{A_{m,n+1}\;B_{n+1}} \over {\rho _m(s)}}\left( {{\delta  \over {\delta x\left( {s+{m \over 2}}
\right)}}} \right)^{n-m}\left\{ {\tau _n\left( {s-{{n-m} \over 2}} \right)\rho _n\left( {s-{{n-m} \over 2}} \right)}
\right\}= \\ &={{A_{m,n+1}\;B_{n+1}} \over {\rho _m\left( {s'-{m \over 2}} \right)}}\left( {{\delta  \over {\delta
x(s')}}} \right)^{n-m}\left\{ {\tau _n\left( {s'-{n \over 2}} \right)\rho _n\left( {s'-{n \over 2}} \right)} \right\}
\end{align*}

With respect to the new variable $s'=s+{m \over 2}$, this expression can be easily differentiated as in (61) giving

\begin{align*} v_{m,n+1}(s)&={{A_{m,n+1}\;B_{n+1}} \over {\rho _m(s)}}\left\{ {\tau _n(s)\left( {{\delta  \over
{\delta x\left( {s+{m \over 2}} \right)}}} \right)^{n-m}\rho _n\left( {s-{{n-m} \over 2}} \right)+}\right. \\
&+\left.{{{\rm sh}(n-m)\omega } \over {{\rm sh}\omega }}\tau '_n\left( {{\delta  \over {\delta x\left( {s+{{m-1}
\over 2}}
\right)}}} \right)^{n-m-1}\rho _n\left( {s-{n \over 2}+{{m-1}
\over 2}} \right) \right\} 
\end{align*}

From (60) we get
\begin{align*} 
{{\nabla v_{m,n}(s)} \over {\nabla x(s)}}&={{\Delta v_{m,n}(s-1)} \over {\Delta
x(s-1)}}=v_{m+1,n}(s-1)=\\ &={{A_{m+1,n\;}B_n} \over {\rho _{m+1}(s-1)}}\left( {{\delta  \over {\delta x\left(
{s+{{m-1} \over 2}} \right)}}} \right)^{n-m-1}\rho _n\left( {s-{n \over 2}+{{m-1} \over 2}} \right)
\end{align*}

Using this result and the values for $A_{m,n}$ given in (58) we get the raising operator for $v_{mn}(s)$, namely,

\begin{align}
v_{m,n+1}(s)=&{B_{n+1} \over B_n}\left\{ {{{\left[ {n+1} \right]} \over {\left[ {n+1-m}
\right]}}{{\lambda _{n+m}} \over {\left[ {n+m} \right]}}{{\left[ n \right]} \over {\lambda _n}}\tau
_n(x)v_{mn}(x) }\right. \nonumber \\
&-\left. {{\left[ {n+1} \right]} \over {\left[ {n+1-m} \right]}}{{\left[ n \right]} \over {\lambda _n}}\tau
'_n\sigma (x){{\nabla v_{m,n}(s)} \over {\nabla x(s)}} \right\}
\end{align}

with $\left[ n \right]\equiv {{shn\omega } \over {sh\omega }}$ corresponding to all values of lattice
functions $x(s)$ given in (50) (51).

In order to construct the lowering operator we use the recurrence relation
\begin{equation} x_m(s)v_{mn}(s)=\tilde \alpha _nv_{m,n+1}(s)+\tilde \beta _nv_{mn}(s)+\tilde \gamma _nv_{m,n-1}(s)
\end{equation} where  $x_m(s)=x\left( {s+{m \over 2}} \right)$ and $v_{mn}(s)\equiv \Delta ^{(m)}y_n(s)$

We introduce the expansion $y_n(s)=a_nx^n(s)+b_nx^{n-1}(s)+\ldots$ in the recurrence relation (66). We have two
cases [17]

\begin{enumerate}
\item [a)] Quadratic lattice: $x(s)=s(s+1)$.
\begin{align*}
 \Delta ^{(m)}x^n(s)&=n(n-1)\ldots (n-m+1)x_m^{n-m}(s)+\\ &+{1 \over {12^m}}n(n-1)\ldots (n-m)(2n-2m+1)x_m^{n-m-1}(s)
\end{align*} which after substitution in the recurrence relation (66) gives
\begin{align}
\tilde \alpha _n&={{a_n} \over {a_{n+1}}}\;{{n-m+1} \over {n+1}}\\
\tilde \beta _n&={{b_n} \over {a_n}}\;{{n-m} \over n}-{{b_{n+1}} \over {a_{n+1}}}\;{{n-m+1} \over {n+1}}-{3 \over
{12^m}}\\
\tilde \gamma _n&={{a_{n-1}} \over {a_n}}\;{{n-m} \over n}\;{{d_{mn}^2} \over {d_{m,n-1}^2}}
\end{align}
\item [b)] Exponential lattice $x(s)=Aq^s+Bq^{-s}$
$$\Delta ^{(m)}x^n(s)=[n]\;[n-1]\ldots [n-m+1]\;x_m^{n-m-1}(s)+C\;x_m^{n-m-3}(s)+\ldots $$ which after substitution
in the recurrence relation (66) gives
\begin{align}
\tilde \alpha _n&={{a_n} \over {a_{n+1}}}\;{{[n-m+1]} \over {[n+1]}}\\
\tilde \beta _n&={{b_n} \over {a_n}}\;{{[n-m]} \over {[n]}}-{{b_{n+1}} \over {a_{n+1}}}\;{{[n-m+1]} \over {[n+1]}}\\
\tilde \gamma _n&={{a_{n-1}} \over {a_n}}\;{{[n-m]} \over {[n]}}\;{{d_{mn}^2} \over {d_{m,n-1}^2}}
\end{align}
\end{enumerate}

Since $a_n=B_n\prod\limits_{k=0}^{n-1} {\left( {-{{\lambda _{n+k}} \over {[n+k]}}} \right)}$ we obtain
\begin{equation}
\tilde \alpha _n=-{{B_n} \over {B_{n+1}}}\;{{\lambda _n} \over {[n]}}\;{{[2n]} \over {\lambda _{2n}}}\;{{[2n+1]}
\over {\lambda _{2n+1}}}\;{{[n-m+1]} \over {[n+1]}},
\end{equation} which after substituting in (65) gives
\begin{equation}
\tilde \alpha _n\;{{\lambda _{2n}} \over {[2n]}}\;v_{m,n+1}(s)={{\lambda _{n+m}} \over {[n+m]}}\;{{\tau _n(s)} \over
{\tau '_n}}v_{mn}(s)-\sigma (s){{\nabla v_{mn}(s)} \over {\nabla x(s)}}
\end{equation}

Inserting the recurrence relation (66) in (74) we get
\begin{equation}
\tilde \gamma _n\;{{\lambda _{2n}} \over {[2n]}}\;v_{m,n-1}(s)=\left\{ {-{{\lambda _{n+m}} \over {[n+m]}}\;{{\tau
_n(s)} \over {\tau '_n}}+{{\lambda _{2n}} \over {[2n]}}}
\right\}v_{mn}(s)+\sigma (s)\;{{\nabla v_{mn}(s)} \over {\nabla x(s)}}
\end{equation}

The last two equation can be considered the raising and lowering operators of generalized orthogonal polynomials on
non-uniform latices for the functions (50) and (51). In the first case the parameter $[n]$ should be taken as $n$.

In order to complete the picture, we define an orthonormal function
\begin{equation}
\Omega _{mn}(s)=d_{mn}^{-1}\sqrt {\rho _m(s)}\;v_{mn}(s)
\end{equation}

Using the properties of the difference operator and the identity
\begin{equation} {{\nabla \rho _m(s)} \over {\rho _m(s)}}={{\tau _{m-1}(s)\;\Delta x_{m-1}\left( {s-{1 \over 2}}
\right)} \over {\sigma (s)+\tau _{m-1}(s)\;\Delta x_{m-1}\left( {s-{1 \over 2}} \right)}}
\end{equation} we get
\begin{align}
\nonumber \nabla \Omega _{mn}(s)&=\sqrt {{{\sigma (s)} \over {\sigma (s)+\tau _{m-1}(s)\;\Delta x_{m-1}\left( {s-{1
\over 2}} \right)}}}\;d_{mn}^{-1}\sqrt {\rho _m(s)}\;\nabla v_{mn}(s)+\\
\nonumber &+{1 \over {\sqrt {\sigma (s)}+\sqrt {\sigma (s)+\tau _{m-1}(s)\;\Delta x_{m-1}\left( {s-{1 \over 2}}
\right)}}} \times \\ & \times {{\tau _{m-1}(s)\;\Delta x_{m-1}\left( {s-{1 \over 2}}
\right)}
\over {\sqrt {\sigma (s)+\tau _{m-1}(s)\;\Delta x_{m-1}\left( {s-{1 \over 2}} \right)}}}\;\Omega _{mn}(s)
\end{align}

Multiplying both sides of (74) by $d_{mn}^{-1}\sqrt {\rho _m(s)}$ and subtituting the value $d_{mn}^{-1}\sqrt {\rho
_m(s)}$ $\nabla v_{mn}(s)$ obtained in (78) we get
\begin{align}
\nonumber &\tilde \alpha _n\;{{\lambda _{2n}} \over {2n}}\;{{d_{m,n+1}} \over {d_{mn}}}\;\Omega _{m,n+1}(s)=
L^+(s,n)\Omega _{mn}(s)=\\
\nonumber &=\left\{ {{{\lambda _{n+m}} \over {[n+m]}}\;{{\tau _n(s)} \over {\tau '_n(s)}}+{{\sqrt
 {\sigma (s)}\;\tau _{m-1}(s)} \over {\sqrt {\sigma (s)}+\sqrt {\sigma (s)+\tau _{m-1}(s)\;\Delta x_m\left( {s-{1
\over 2}} \right)}}}{{\nabla x_m\left( {s+{1 \over 2}} \right)}
\over {\nabla x(s)}}} \right\}\Omega _{mn}(s)-\\ &-\sqrt {\sigma (s) {\sigma (s)} +\tau _{m-1}(s)\;\Delta
x_{m-1}\left( {s-{1 \over 2}} \right)}{{\nabla \Omega _{mn}(s)} \over {\nabla x(s)}}
\end{align}

Similarly
\begin{align}
\nonumber &\tilde \gamma _n\;{{\lambda _{2n}} \over {2n}}\;{{d_{m,n-1}} \over {d_{mn}}}\;\Omega _{m,n-1}(s)=
L^-(s,n)\Omega _{mn}(s)=\\
\nonumber &=\left\{ {-{{\lambda _{n+m}} \over {[n+m]}}\;{{\tau _n(s)} \over {\tau '_n}}+{{\lambda _{2n}}
\over {2n}}\left( {s-\tilde \beta _n} \right)-{{\sqrt {\sigma (s)}\;\tau _{m-1}(s)} \over {\sqrt {\sigma (s)+\tau
_{m-1}(s)\;\Delta x_m\left( {s-{1 \over 2}}
\right)}}}{{\nabla x_m\left( {s+{1 \over 2}} \right)} \over {\nabla x(s)}}} \right\}\times \\ &\qquad \Omega _{mn}(s)
+\sqrt {\sigma (s){\sigma (s)}+\tau _{m-1}(s)\;\Delta x_{m-1}\left( {s-{1
\over 2}} \right)}{{\nabla \Omega _{mn}(s)} \over {\nabla x(s)}}
\end{align}

The last two expressions can be considered the raising and lowering operators for the generalized orthonormal
functions on non-homogeneous lattices of the type (50) and (51). It can be proved that these operators are mutually
adjoint with respect to the scalar product of unit weight.

As in the previous sections we can factorize the raising and lowering operators as follows:
$$L^-(s,n+1)L^+(s,n)=\mu (n)+u(s+1,n)H(s,n)$$
$$L^+(s,n)L^-(s,n+1)=\mu (n)+u(s,n-1)H(s,n+1)$$
where
$$\mu (n)={{\lambda _{2n}} \over {\left[ {2n} \right]}}\,{{\lambda _{2n+2}} \over {\left[ {2n+2} \right]}}\,\tilde
\alpha _n\,\tilde \gamma _{n+1},$$
$$u (s,n)={{\lambda _n} \over {\left[ n \right]}}\,{{\tau _n(s)} \over {\tau '_n}}-{{\sigma (s)} \over {\nabla
x(s)}}$$
and $H(s,n)$ is the difference operator derived from the left side of (47) after substituting $\Omega _{mn}(s)$
instead $v _{mn}(s)$ given in (76). Notice that the expresions for the factorization of the raising and lowering
operators becomes the same expresions (32) and (33) given in [15].

\section{Conclusions}

We have developped the construction of raising and lowering operators for classical OP of discrete variable on
non-homogeneous lattice extended also to the generalized OP on homogeneous and non-homogeneous lattice.

In the last case (generalized OP) the raising and lowering operators can be defined with respect to the index $n$,
the order of the OP, or with respect to the index $m$, the order of the difference derivative of the generalized OP,
or both.

In our work we have taken into account only the index $n$, although we have suggest how to complement the calculus
with the index $m$. We have also introduced the orthornomal functions of unit weight, more suitable to quantum
mechanical applications.

Our presentation leads to an easier way for the continuous limit (compair with a different presentation in [5]).

We have already worked out some physical application of raising and lowering operators on homogeneous and
non-homogeneous lattice. For instance, the quantum mechanical models for the harmonic oscillator in one dimension
(Kravchuk OP), the hydrogen atom (generalized Meixner OP) [22], the Heisenberg equation of motion on the lattice
(Hahn OP) [23] Dirac and Klein-Gordon equation on a homogeneous lattice (discrete exponential function). [24] [25]

Finally the connection between OP on non-homogeneous lattice and the 3nj-Wigner coefficients and its application to
spin networks models in quantum gravity are now in progress.

\bigskip
{\noindent \bf \large Acknowledgments}
\bigskip

The autor wants to express his gratitude to the referees for their suggestions and new references. This work has
been partially supported by Ministerio de Ciencia y Tecnolog{\'\i}a, grant BFM2000-0357.


\begin{thebibliography}{00}
\bibitem{} A.F.Nikiforov, S.K.Suslov, V.B.Uvarov, {\em Classical orthogonal polynomials of a discrete variable},
Springer, Berlin 1991.
\bibitem{} Although some authors call these polynomials associated OP, we prefer
to  call them generalized OP in order to distinguish from the traditional
name of associated classical OP. See Mizan Rahman, ``The
associated classical orthogonal polynomials'', in
Special Functions 2000 (J. Bustoz et al. eds.) Kluwer
Academic Publishers, Netherlands 2001.
\bibitem{} Ref. 1, page 8.
\bibitem{} Ref. 1, page 6.
\bibitem{} M. Lorente, ``Raising and lowering opertors, factorization and differential/difference operator of
hypergeometric type'', {\em J. Phys. A: Math. Gen. 34} (2001) 569-588.
\bibitem{} Ref. 1, page 11.
\bibitem{} Ref. 1, page 19.
\bibitem{} Ref. 1, page 20.
\bibitem{} Ref. 1, page 24.
\bibitem{} Ref. 1, page 26.
\bibitem{} Ref. 1, page 42-46.
\bibitem{} Ref. 1, page 55.
\bibitem{} N.M. Atakishiyev, S.K. Suslov, ``About on class of special functions'', {\em Rev. Mex. Fis. 34} (1988)
152-167.
\bibitem{} N.M. Atakishiyev, S.K. Suslov, ``On classical orthogonal polynomials'', {\em Constr. Approx. 11} (1995)
181-226.
\bibitem{} R. Alvarez-Nodarse, R.S. Costas-Santos, ``Factorization method for difference equation of hypergeometric
type on non-uniform lattice'' {\em J. Phys. A: Math. Gen. 34} (2001) 555.
\bibitem{} R. Alvarez-Nodarse, J. Arves\'u, ``On the q-polynomials in the exponential lattice'', {\em Integral
transforms and special function 8} (1999) 299-324.
\bibitem{} Ref. 1, page 101-104.
\bibitem{} A.F. Nikiforov, V.B. Uvarov, ``Polynomial solutions of hypergeometric type, difference equations and their
classification'', {\em Integral transforms and special functions 1} (1993) 223-249.
\bibitem{} Y.F. Smirnov, ``On factorization and Algebraization of difference equation of hypergeometric type''. {\em
Proc. Int. Workshop on Orthogonal Polynomial in Math. Physics}, ed. M. Alfaro et al., Universidad Carlos III,
Legan\'es (Madrid) 1997, p. 153.
\bibitem{} Y.F. Smirnov, ``Finite difference equation and factorization method'', {\em Proc. V Wigner Symposium},
ed. P. Kasperkovitz and D. Grano, Singapore, World Scientific (1998) p. 148.
\bibitem{} Y.F. Smirnov, ``Factorization method: new aspects''. {\em Rev. Mex. Fis. 45} (suppl. 2) (1999) 1-6.
\bibitem{} M. Lorente, ``Continuous vs discrete models for the quantum harmonic oscillator and the hydrogen atom'',
{\em Phys. Lett, A 285} (2001) 119-126.
\bibitem{} M. Lorente, ``On some integrable one-dimensional quantum mechanical systems'', {\em Phys, Lett, B 223}
(1989) 345-350.
\bibitem{} M. Lorente, ``A new scheme for the Klein-Gordon and Dirac field on the lattice with axial anomaly'', {\em
J. Group Theory in Phys. 1} (1993) 105-121.
\bibitem{} M. Lorente, P. Kramer, ``Representations of the discrete inhomogeneous Lorentz group and Dirac wave
equation on the lattice'', {\em J. Phys. A: Math. Gen. 32} (1999) 2481-2497.

\end{thebibliography}
\end{document}